\documentclass[aps,english,preprintnumbers,nofootinbib,twocolumn]{revtex4-1}

\usepackage{amsfonts,amsmath,amssymb}
\usepackage{graphicx}
\usepackage[utf8]{inputenc}
\usepackage{hyperref}
\usepackage{babel}
\usepackage{xcolor}  

\newcommand\be{\begin{equation}}
\newcommand\ee{\end{equation}}
\newcommand\bea{\begin{eqnarray}}
\newcommand\eea{\end{eqnarray}}
\usepackage{tikz}
\usepackage{tikz-3dplot}

\begin{document}

\title{Timelike Holographic Complexity}
\author{Mohsen Alishahiha}
\affiliation{ School of Quantum Physics and Matter\\ Institute for Research in Fundamental Sciences (IPM),\\
	P.O. Box 19395-5531, Tehran, Iran}

\begin{abstract}
Motivated by the pseudo-entropy program, we study timelike subregion complexity within the holographic Complexity=Volume framework, extending previous spatial constructions to Lorentzian boundary intervals. For hyperbolic timelike regions in pure AdS, we compute the enclosed bulk volume and show that, despite the Lorentzian embedding, the resulting complexity is purely real under the stated volume prescription. We generalize the analysis to AdS black brane geometries, where the extremal-surface configurations considered may either remain outside the horizon or penetrate it, with the horizon-to-turning-point segment being timelike in the exterior case and spacelike in the interior case. In both configurations, the complexity exhibits the same leading UV-divergence structure as the spacelike case and receives no imaginary contribution when the horizon terms are combined with a common regulator. This reality contrasts with the potentially complex-valued pseudo-entropy and reflects the chosen Lorentzian volume prescription, while a precise boundary interpretation of the proposed complexity remains conjectural.
\end{abstract}
\maketitle

\section{Introduction}

Quantum entanglement has long been recognized as a fundamental probe of quantum correlations and spacetime structure in holographic theories \cite{Ryu:2006bv,Ryu:2006ef,Hubeny:2007xt}. Traditionally, entanglement entropy is defined for spatial subregions on a fixed time slice, capturing the pattern of quantum correlations among spatially separated degrees of freedom in the boundary theory. More recently, this framework has been broadened to quantities associated with transitions between quantum states, notably pseudo-entropy
\cite{Nakata:2020luh,Mollabashi:2020yie,Mollabashi:2021xsd}. Related developments extend entanglement-like measures to timelike subregions, thereby enlarging the scope of quantum information probes available in holographic systems.

Pseudo-entropy generalizes entanglement entropy to situations involving nontrivial quantum state transitions. Given nonorthogonal initial and final states $|\psi_i\rangle$ and $|\psi_f\rangle$, one defines a transition matrix
\begin{equation}\label{TM}
    \tau = \frac{|\psi_f\rangle \langle \psi_i|}{\langle \psi_i | \psi_f \rangle},
\end{equation}
and the pseudo-entropy of a subsystem $A$ (for a bipartite system $A \cup B$) as
\begin{equation}
    S_A^{\mathrm{pseudo}} = - \mathrm{Tr}\, \tau_A \log \tau_A,
    \qquad \tau_A = \mathrm{Tr}_B\, \tau .
\end{equation}
Unlike conventional entanglement entropy, pseudo-entropy can be complex because the reduced transition matrix $\tau_A$ need not be Hermitian. Its real and imaginary parts contain information about the chosen state transition, although their interpretation depends on the states and the construction. The emergence of complex values thus reflects the richer analytic structure of these generalized information measures.

Within the AdS/CFT correspondence \cite{Maldacena:1997re}, pseudo-entropy naturally motivates the study of extremal-surface configurations anchored on timelike boundary intervals \cite{Doi:2022iyj,Doi:2023zaf,Narayan:2022afv}. These constructions extend the use of the Ryu--Takayanagi (RT) prescription
\cite{Ryu:2006bv,Ryu:2006ef} to timelike anchoring regions and may involve both spacelike and timelike surface segments. Such generalizations have attracted increasing attention in recent years (see, \textit{e.g.}, \cite{Wang:2018jva,Narayan:2020nsc,Liu:2022ugc,Li:2022tsv,Narayan:2023ebn,Narayan:2023zen,Jena:2024tly,
Afrasiar:2024lsi,Afrasiar:2024ldn,Nunez:2025gxq,Zhao:2025zgm,Jiang:2025pen,Mohan:2025aiw}). In these constructions, the corresponding holographic entanglement entropy for timelike subregions typically acquires an imaginary part, connecting timelike entanglement entropy with complex-valued pseudo-entropy.

This naturally raises the question of whether other quantum information measures---most notably quantum complexity---admit a consistent timelike generalization, and if so, what boundary quantity they compute.

Holographic complexity has been proposed as a measure of the computational cost required to prepare a given boundary state from a simple reference state using a minimal set of quantum operations. Two prominent proposals are the ``Complexity=Volume'' (CV) conjecture \cite{Susskind:2014rva,Stanford:2014jda}, which was further extended to spatial subregion complexity in \cite{Alishahiha:2015rta} (see also \cite{Ben-Ami:2016qex,Abt:2017pmf}), and the ``Complexity=Action'' (CA) conjecture \cite{Brown:2015bva,Brown:2015lvg}, whose subregion version has been explored in \cite{Carmi:2016wjl,Agon:2018zso,Alishahiha:2018lfv}. These frameworks have substantially deepened our understanding of information storage, state preparation, and dynamical complexity growth in holographic field theories, primarily for spatially defined subsystems and global states.

Motivated by the pseudo-entropy program, in this work we introduce and analyze a timelike version of subregion complexity within the CV proposal. This involves evaluating the bulk volume bounded by the extremal-surface configurations associated with a timelike boundary region, thereby extending the spatial construction of \cite{Alishahiha:2015rta} to timelike anchoring data. We regard the resulting geometric quantity as a candidate for a notion of complexity associated with finite timelike intervals or transitions between boundary configurations. A precise identification with a boundary circuit or transition complexity remains conjectural.

A crucial feature of this construction is that the bulk volume is evaluated over a real Lorentzian region using the stated volume prescription. For the black-brane configurations, the horizon contributions are combined with a common regulator. The resulting quantity is real under these prescriptions, whereas timelike entanglement entropy can be complex. This difference reflects the geometric definition of the proposed volume. 
In particular, a linear dependence of the finite part on $T$, where present, describes interval-size dependence and is not a dynamical complexity growth rate. Such a relation does not by itself establish an identification with circuit complexity.

We first focus on timelike hyperbolic subregions in AdS$_{d+2}$, building on the explicit extremal surface solutions constructed in \cite{Doi:2022iyj,Doi:2023zaf}. We compute the associated bulk volumes and analyze their divergence structure in detail. We find that timelike subregion complexity remains purely real under the stated prescription, providing a geometric quantity complementary to timelike entanglement-based measures.

We then extend our analysis to thermal states dual to AdS black brane geometries, where the presence of an event horizon introduces qualitatively new features. In this setting, the configurations considered have turning points either outside or inside the horizon. Their causal character is determined by the induced metric: the horizon-to-turning-point segment is timelike for an exterior turning point and spacelike for an interior turning point. The resulting complexity exhibits universal UV divergences together with finite, horizon-sensitive contributions and remains real under the prescribed regularization. For $d>1$, we identify a dimension-dependent bound on the first accessible turning point of the real interior family. The exterior relation between interval length and turning point is nonmonotonic, so a fixed interval need not select a unique configuration. Global matching and admissibility conditions remain necessary for a complete boundary interpretation.

The remainder of this letter is organized as follows. In Sec.~II, we compute timelike subregion complexity for hyperbolic regions in pure AdS and discuss its general properties. In Sec.~III, we generalize the construction to AdS black brane backgrounds and analyze the dependence of the extremal surfaces and complexity on interval length near and beyond the horizon. We conclude in Sec.~IV with a summary of our results and a discussion of future directions.



\section{Timelike Subregion and Holographic Volume}

In standard holographic constructions, a 
subregion is defined as a spatial domain on 
a constant-time slice. In contrast, here we
consider timelike subregions, namely 
codimension-one regions that extend along the 
time direction while being localized in space.
Such regions have appeared in the study of
timelike entanglement entropy and pseudo-entropy.
These objects should not be interpreted as
subsystems in the usual quantum information
sense, since they do not admit a tensor 
factorization of the Hilbert space. Rather, 
they encode information about the evolution of 
a spatial region over a finite time interval.

Following \cite{Doi:2022iyj,Doi:2023zaf}, we consider a timelike boundary subregion in an AdS$_{d+2}$ spacetime with $d \ge 1$. The bulk geometry is described by the metric
\begin{equation}
    ds^2 = \frac{R^2}{r^2}\left(dr^2 + dy^2 - d\xi^2 + \xi^2 dX_{d-1}^2\right),
\end{equation}
where $y$ is a spectator coordinate, 
$X_{d-1}$ denotes the
$(d-1)$-dimensional hyperbolic space 
(in pure AdS), and $R$ is the AdS radius. 
The boundary subsystem is chosen to be a 
timelike hyperbolic region defined by
$\xi^2 \le \frac{T^2}{4}$, with $T$ the length of the
boundary time interval. Varying this length compares boundary
regions of different temporal extent; it does not describe the
time evolution of a fixed boundary configuration. Accordingly,
small and large $T$ refer to small and large intervals.

The area functional for a codimension-two extremal surface anchored on this region is
\begin{equation}
    A = H_{d-1} R^d \int dr\, \frac{\xi^{d-1}\sqrt{1-\xi'^2}}{r^d},
\end{equation}
where $H_{d-1} = \mathrm{Vol}(X_{d-1})$ is 
the transverse volume. Extremizing this
functional yields two distinct branches 
of solutions \cite{Doi:2022iyj,Doi:2023zaf},
\begin{equation}\label{eq:branches}
   \xi^2 = r^2 + \frac{T^2}{4}, \qquad 
   \xi^2 = r^2 - \frac{T^2}{4},
\end{equation}
corresponding to spacelike and timelike extremal surfaces, respectively. The timelike branch exists only for $r \ge T/2$, thereby introducing a causal cutoff in the bulk geometry.

Together, these two branches form the
complete extremal surface configuration
associated with the timelike subregion. 
As shown in \cite{Doi:2022iyj,Doi:2023zaf},
while the area of the spacelike branch remains
real, the timelike branch contributes an 
imaginary term to the holographic entanglement
entropy. This feature motivates examining the 
corresponding construction in holographic 
complexity, where one might expect qualitatively
different behavior due to the volume-based 
nature of the proposal.

In this work we propose a geometric quantity
associated with a timelike boundary region and 
conjecture that it captures a notion of 
complexity in the dual field theory. Since 
the boundary region extends along a timelike
direction, it does not define a subsystem in
the usual sense and therefore does not admit
a standard reduced density matrix description.
Instead, we adopt an interpretation inspired by
pseudo-entropy and timelike entanglement entropy.
Consider a spatial subregion $A$ of the boundary
theory and two Cauchy slices at times $t_1$ and $t_2$.
One may associate to these slices the reduced states
$\rho_A(t_1)$ and $\rho_A(t_2)$, or more generally a
transition matrix between them. We propose that the 
bulk quantity defined in this paper provides a
measure of the complexity associated with transforming
the configuration of the system in region $A$ at 
time $t_1$ to that at time $t_2$. In this sense, 
it may be viewed as a candidate for a transition 
complexity associated with a timelike interval.

A precise definition of the boundary state associated 
with a timelike region requires specifying the algebra 
of observables localized on such a region. This is subtle,
as timelike-separated regions do not define commuting 
operator algebras. In analogy with pseudo-entropy, one may
instead consider transition amplitudes or generalized 
density matrices associated with pairs of states. 
In this work, we do not attempt to construct this object
explicitly. Rather, we take a pragmatic approach and 
propose that the bulk geometric quantity captures a 
complexity-like measure associated with the transition
between configurations of the system in a fixed spatial
region at different times. At present, this identification
should be regarded as a conjecture. A first-principles
derivation from the boundary theory, including a precise 
definition of the relevant operator algebra associated 
with a timelike region, remains an open problem. 
Nevertheless, the geometric construction we propose 
satisfies several properties expected of a complexity 
measure, such as positivity, real-valuedness, and a 
natural dependence on the length of the timelike interval.

It is natural to expect that such a transition complexity
may be related to existing field-theoretic notions of 
quantum complexity, such as circuit complexity defined
via unitary gates, path-integral optimization, or
Lorentzian tensor network constructions. In these
approaches, one attempts to quantify the minimal 
computational cost required to transform one quantum 
state into another under suitable locality and causality 
constraints. While a precise mapping between our 
geometric proposal and these frameworks remains to be 
established, the qualitative features we observe—such
as positivity and real-valuedness—are consistent with expectations
from these definitions. Any linear dependence on interval length,
where present, describes dependence on the size of the timelike
region and is not a dynamical complexity growth rate. It therefore
does not establish linear growth of circuit complexity under time
evolution. We leave a more direct field-theoretic
formulation of timelike subregion complexity for future work.

Within the subregion Complexity=Volume (CV) prescription
\cite{Alishahiha:2015rta}, holographic complexity is
proportional to the bulk volume enclosed by the relevant 
extremal surfaces,
\begin{equation}\label{eq:V_def}
   V = H_{d-1} R^{d+1} \int_{\xi \le f(r)} dr\, d\xi\; \frac{\xi^{d-1}}{r^{d+1}},
\end{equation}
where $f(r)$ denotes the extremal profiles given in 
Eq.~\eqref{eq:branches}. The bulk region entering our definition
is constructed geometrically as the domain bounded by the boundary timelike region and the associated extremal surfacess, as illustrated in Fig~\ref{fig:C1}. Since the extremal surfaces are defined covariantly, the resulting construction is independent of coordinate choices. In all examples considered in this work, the symmetry of the setup uniquely determines the bulk region.

Performing the $\xi$ integration yields the volume difference between the two branches,
\begin{equation}\label{eq:V_difference}
 V = \frac{2H_{d-1} R^{d+1}}{d}\!\left[ 
 \int_{\epsilon}^{\infty}\! dr\,
 \frac{\bigl(r^2+\frac{T^2}{4}\bigr)^{\frac{d}{2}}}{r^{d+1}} 
 - \!\!\int_{\frac{T}{2}}^{\infty}\! dr\,
 \frac{\bigl(r^2 - \frac{T^2}{4}\bigr)^{\frac{d}{2}}}{r^{d+1}} 
 \right],
\end{equation}
where $\epsilon$ is a UV cutoff near the AdS boundary at $r=0$. The lower limit of the second integral reflects the fact that the timelike branch exists only for $r \ge T/2$, while the overall factor of 2 arises from symmetry of the configuration.

\begin{figure}[ht]
\centering
\begin{tikzpicture}[scale=2]
\coordinate (O) at (0,0);
\coordinate (O1) at (0,-1);
\coordinate (r) at (1.3,0);
\coordinate (t) at (0,1);
\draw[->] (O) -- (r) node[right] {$r$};
\draw[->] (O1) -- (t) node[above] {$\xi$};
\draw[thick, green] plot[domain=0:1, samples=100] (\x, {sqrt(1/16+\x^2)});
\draw[thick, green] plot[domain=0:1, samples=100] 
(\x, {-sqrt(1/16+\x^2)},0);
\draw[thick, red] plot[domain=1/4:1, samples=100] (\x, {sqrt(-1/16+\x^2)});
\draw[thick, red] plot[domain=1/4:1, samples=100] 
(\x, {-sqrt(-1/16+\x^2)});
\node at (-0.13, 1/4, 0) {$\frac{T}{2}$};
\node at (-0.17, -1/4, 0) {$-\frac{T}{2}$};
\node [below] at ( 1/4+0.1,0) {$\frac{T}{2}$};
\fill[blue!20, opacity=0.5]
  (0,0)
  plot[domain=0:1, samples=100] (\x, {sqrt(1/16+\x^2)})
  -- plot[domain=1:1/4, samples=100] (\x, {sqrt(-1/16+\x^2)})
  -- plot[domain=1/4:1., samples=100] (\x, {-sqrt(-1/16+\x^2)})
  -- plot[domain=1:0, samples=100] (\x, {-sqrt(1/16+\x^2)})
  -- cycle;
\end{tikzpicture}
\caption{The colored region between the spacelike (green) and timelike (red) extremal surfaces represents the bulk volume associated with the timelike subregion complexity.}
\label{fig:C1}
\end{figure}

Following \cite{Alishahiha:2015rta}, the timelike subregion complexity is defined as
\begin{equation}\label{complexity}
   \mathcal{C}_{\mathrm{T}} = \frac{V}{G R},
\end{equation}
where $G$ is the $(d+1)$-dimensional Newton constant.  We regard this construction as a minimal covariant extension of subregion CV complexity to timelike domains, defined by extremal surfaces compatible with boundary causal structure.

It is important to emphasize that although neither 
branch alone defines a conventional homology region, 
the pair together admits a natural covariant 
interpretation: they bound the maximal real Lorentzian
bulk domain that can be reached by spacelike surfaces 
anchored to the timelike boundary interval. The bulk 
volume entering the CV prescription is therefore not chosen \textit{ad hoc}, but is fixed uniquely by causal and geometric considerations.

Although each integral in Eq.~\eqref{eq:V_difference} separately exhibits an infrared divergence as $r \to \infty$, these divergences cancel in the difference, leaving a finite IR contribution. The remaining UV divergences are regulated by the cutoff $\epsilon$.

While the integrals in Eq.~\eqref{eq:V_difference} can be expressed in terms of hypergeometric functions for arbitrary $d$, it is more illuminating to present explicit results in lower dimensions. For $d=1,2,3,4$, we find
\begin{align}\label{eq:exact}
\text{AdS}_3:&\quad {\cal C}_{\rm T}=\frac{R}{G}\,\frac{T}{\epsilon},\nonumber\\
\text{AdS}_4:&\quad {\cal C}_{\rm T}=\frac{V_1R^2}{G}\left(\frac{T^2}{8\epsilon^2}+\log\frac{T}{2\epsilon}+\frac{1}{2}\right),\nonumber\\
\text{AdS}_5:&\quad {\cal C}_{\rm T}=\frac{V_2R^3}{3G}\left(\frac{T^3}{12\epsilon^3}+\frac{3T}{4\epsilon}\right),\nonumber\\
\text{AdS}_6:&\quad {\cal C}_{\rm T}=\frac{V_3R^4}{2G}\left(\frac{T^4}{64\epsilon^4}+\frac{T^2}{4\epsilon^2}+\log\frac{T}{2\epsilon}+\frac{3}{4}\right).
\end{align}

The divergence structure closely parallels that of spacelike subregion complexity: power-law UV divergences associated with short-distance correlations, and universal logarithmic terms appearing in even dimensions. The leading divergence scales with the volume of the boundary subregion, confirming that the dominant contribution to holographic complexity arises from the near-boundary geometry.

A key distinction from timelike entanglement entropy now becomes apparent. The expressions in Eq.~\eqref{eq:exact} are manifestly real. While the timelike branch of the extremal surface contributes an imaginary term to entanglement entropy—due to analytic continuation across the light cone—the complexity computation integrates over a strictly real Lorentzian bulk region with $r \ge T/2$. As a result, the CV prescription yields a real-valued geometric quantity.

The fact that the proposed quantity is real-valued follows directly from the choice of the Lorentzian volume element $\sqrt{-\det g_{\text{ind}}}$, which is the natural generalization of the standard CV prescription to timelike hypersurfaces. This choice ensures that the resulting volume is real and positive, as expected for a complexity measure. We emphasize that this prescription is a direct Lorentzian continuation of the standard CV proposal for spacelike hypersurfaces, in which the volume element is defined using the induced metric on the bulk slice. The use of $\sqrt{-\det g_{\text{ind}}}$ is therefore not arbitrary, but follows from requiring covariance and consistency with the Euclidean/spacelike formulation. Alternative prescriptions would either violate reparametrization invariance or fail to reproduce the standard CV result in the appropriate limit. In contrast, the complex nature of timelike holographic entanglement entropy arises from the use of length functionals that do not incorporate this sign adjustment. From this perspective, the reality of our result is not a dynamical feature but rather a consequence of adopting a physically appropriate definition of volume in Lorentzian signature.

To build further intuition, consider the three-dimensional case, where the extremal surfaces can be visualized explicitly in both Poincar\'e and global coordinates, as shown in Fig.~\ref{fig2} (see also \cite{Doi:2022iyj}). In the entanglement entropy computation, the timelike branch becomes evident upon transforming to global coordinates, giving rise to the imaginary contribution characteristic of pseudo-entropy. For complexity, however, the relevant bulk volume is not that enclosed by the spacelike surface alone, but rather the region bounded jointly by the spacelike and timelike branches, illustrated by the shaded domain in Fig.~\ref{fig2}.

\begin{figure}[ht]
\centering
\begin{tikzpicture}[scale=1.68]
\coordinate (O) at (0,0);
\coordinate (O1) at (0,-1);
\coordinate (r) at (1.5,0);
\coordinate (t) at (0,1);
\coordinate (L1) at (2.5,-1.2);
\coordinate (L11) at (2.5,1.2);
\coordinate (R1) at (3.7,-1.2);
\coordinate (R11) at (3.7,1.2);
\coordinate (ti1) at (3.1,-0.6);
\coordinate (ti2) at (3.1,0.6);
\coordinate (b1) at (3.7,-0.45);
\coordinate (b2) at (3.7,0.45);
\coordinate (tau1) at (2.3,0.8);
\coordinate (tau2) at (2.3,1);
\coordinate (rho1) at (3.2,-1.4);
\coordinate (rho2) at (3.5,-1.4);
\draw[->] (O) -- (r) node[right] {$z$};
\draw[->] (O1) -- (t) node[above] {$t$};
\draw[-] (L1) -- (L11) ;
\draw[-] (R1) -- (R11) ;
\draw[-] (L11) -- (R11) ;
\draw[-] (L1) -- (R1) ;
\draw[-,thick, red] (ti1) -- (ti2) ;
\draw[-,thick, blue] (b1) -- (b2) ;
\draw[->] (tau1) -- (tau2) node[above] {$\tau$};
\draw[->] (rho1) -- (rho2) node[right] {$\rho$};
\draw[thick, green] plot[domain=0:1, samples=100] (\x, {sqrt(1/16+\x^2)});
\draw[thick, blue] plot[domain=-1/4:1/4, samples=100] (0,\x , 0);
\draw[thick, gray] plot[domain=0:1.2, samples=100] (\x+2.5, {\x});
\draw[thick, gray] plot[domain=0:1.2, samples=100] (\x+2.5, {-\x});
\draw[thick, green] plot[domain=0:1, samples=100] 
(\x, {-sqrt(1/16+\x^2)},0);
\draw[thick, green] plot[domain=3.1:3.7, samples=100] 
(\x, {3.1/\x-0.4});
\draw[thick, green] plot[domain=3.1:3.7, samples=100] 
(\x, {-3.1/\x+0.4});
\node at (-0.17, 1/4, 0) {$\frac{T}{2}$};
\node at (-1/4, -1/4, 0) {$-\frac{T}{2}$};
\node at (3.98, 1.2) {$\pi$};
\node at (3.9, -1.2) {$-\pi$};
\node at (3.98, 0.45) {$\frac{T}{2}$};
\node at (3.9, -0.45) {$-\frac{T}{2}$};
\node at (3.9, -0.45) {$-\frac{T}{2}$};
\node at (2.3, 0) {$0$};
\shade[top color=blue!10, bottom color=blue!5, opacity=0.4]
  plot[domain=3.1:3.7, samples=100] (\x, {3.1/\x-0.4}) --
  plot[domain=3.7:3.1, samples=100] (\x, {-3.1/\x+0.4}) -- cycle;
\end{tikzpicture}
\caption{Extremal surfaces for a timelike entangling region (blue interval) in Poincaré (left) and global (right) coordinates. In the global case, in addition to the spacelike extremal surface (green), there exists a timelike one (red). The timelike subregion complexity corresponds to the bulk volume enclosed between these two surfaces, illustrated by the shaded (colored) region. This figure essentially reproduces Fig.~3 of \cite{Doi:2022iyj} for clarity and comparison. Here, $(r, z)$ and $(\tau, \rho)$ denote the Poincar\'e and global coordinate systems, respectively.}
\label{fig2}
\end{figure}

One might wonder whether the coexistence of spacelike and timelike branches implies a hidden causal connection that could, in principle, generate an imaginary contribution to the complexity. At present, however, it remains unclear whether such a causal volume genuinely exists \footnote{I would like to thank T.~Takayanagi for a comment on this point.}. Instead, the explicit calculation shows that any potential phase factors cancel, leaving a purely real result. Indeed, from a technical standpoint, for odd $d$ both endpoints of the timelike integral acquire phase factors that precisely cancel. Timelike subregion complexity therefore measures only the real, causally accessible Lorentzian bulk volume, while pseudo-entropy captures additional phase information associated with temporal correlations.

Comparing with the spacelike subregion results of \cite{Alishahiha:2015rta}, we find that although the divergence structure of timelike complexity closely mirrors the spacelike case, the two are not related by a simple Wick rotation. Both exhibit identical hierarchies of UV divergences, but their finite terms differ both quantitatively and conceptually. In particular, for odd-dimensional AdS backgrounds, timelike complexity lacks the universal constant term present in the spacelike case. This absence suggests that timelike complexity has a genuinely distinct geometric origin, rooted in the Lorentzian causal structure of the extremal hypersurface.

This distinction can be further clarified by reconsidering the extremization problem for a spacelike spherical subregion at the boundary. The corresponding area functional admits two solutions,
\begin{equation}
   \xi(r) = \sqrt{\ell^2 - r^2}, \qquad 
   \xi(r) = i\sqrt{\ell^2 + r^2},
\end{equation}
where $\ell$ is the radius of the boundary region. The first solution is the familiar Ryu--Takayanagi surface \cite{Ryu:2006bv}, which underlies standard computations of entanglement entropy and spacelike subregion complexity \cite{Alishahiha:2015rta}. The second solution is purely imaginary and is usually discarded as unphysical, since it does not correspond to a real Lorentzian surface in the bulk. From the present perspective, however, this ``discarded'' branch plays a crucial role. Upon analytic continuation, it reproduces precisely the timelike extremal branch encountered in timelike entanglement entropy and timelike subregion complexity. The imaginary embedding $\xi = i\sqrt{\ell^2 + r^2}$ represents a continuation across the light cone, encoding the causal extension of the boundary region into the timelike domain. Its contribution is responsible for the imaginary part of pseudo-entropy, reflecting phase information associated with non-unitary or time-reflected evolutions of the boundary state.

In this sense, the real and imaginary branches are not independent, but rather analytic continuations of a single underlying geometric structure. The real branch probes spatial entanglement, while the imaginary branch captures temporal and causal correlations. This unified picture naturally bridges spacelike and timelike holographic observables \footnote{It is worth noting that \cite{Heller:2024whi,Heller:2025kvp} propose a systematic holographic framework for timelike entanglement entropy based on complexified bulk geometries. I thank M.~Heller for bringing these insightful works to my attention.}.

Within the CV framework, however, only the real Lorentzian bulk volume contributes directly, yielding a strictly real measure of complexity. Although the extremal configuration for timelike subregions includes a timelike branch, the resulting complexity remains real, underscoring the fundamentally geometric—as opposed to entropic—nature of holographic complexity.



\section{Timelike Subregion Complexity for AdS Black Branes}
\label{sec:blackhole}

It is natural to extend our discussion of timelike subregion complexity to thermal states, which are holographically dual to AdS black hole or black brane geometries. In this context, the thermal nature of the boundary state is encoded in the presence of an event horizon, which introduces an additional geometric scale and modifies the causal structure relevant for extremal surfaces anchored to timelike subregions. For concreteness, we consider a $(d+2)$-dimensional AdS black hole spacetime with metric
\begin{equation}\label{BB}
    ds^2 = \frac{R^2}{r^2}\left(-f(r)\,dt^2 + \frac{dr^2}{f(r)} + dx^2 + dY_{d-1}^2\right),
\end{equation}
where
\begin{equation}
    f(r) = 1 - \frac{r^{d+1}}{r_h^{d+1}},
\end{equation}
and $r_h$ denotes the horizon radius. The conformal boundary is located at $r=0$. Here, $R$ is the AdS radius, and $dY_{d-1}^2$ denotes the metric on the transverse $(d-1)$-dimensional space. The temperature of the dual thermal state is given by the Hawking temperature $T_H = (d+1)/(4\pi r_h)$.

We consider a timelike subregion on the boundary defined by a finite time interval $t\in[-T/2,T/2]$ at fixed spatial coordinate $x$, while extending uniformly along the transverse $y_i$ directions. Although this configuration lacks the manifest symmetry of the hyperbolic subregions discussed earlier, the extremization procedure can still be carried out by considering embeddings of the form $t=t(r)$, following the general framework for timelike holographic complexity \cite{Doi:2023zaf}. The induced metric on such a hypersurface is
\begin{equation}
    ds_{\text{ind}}^2 = \frac{R^2}{r^2}\left[\left(\frac{1}{f(r)} - f(r)\, t'(r)^2\right)dr^2 + dY_{d-1}^2\right],
\end{equation}
where $t'(r)=dt/dr$. The corresponding area functional reads
\begin{equation}
    A = L^{d-1} R^d \int dr\, \frac{\sqrt{\frac{1}{f(r)} - f(r)\,t'(r)^2}}{r^d},
\end{equation}
with $L^{d-1} = \mathrm{Vol}(Y_{d-1})$. Extremizing this functional yields a conserved quantity,
\begin{equation}
    \frac{-f(r)t'(r)}{r^d \sqrt{\frac{1}{f(r)} - f(r)t'(r)^2}} = \text{constant},
\end{equation}
which is the conserved momentum conjugate to $t$. This conservation law follows from the fact that the Lagrangian does not depend explicitly on $t$. For the area functional written above, this constant is real on spacelike segments and imaginary on timelike segments; a real-valued $t'(r)$ does not by itself fix the reality of the conserved momentum. The causal character of each segment must therefore be determined from the induced metric, independently of whether it lies outside or inside the horizon.

To systematically capture the different possible extremal surface configurations anchored to timelike intervals, it is convenient to square the conservation equation and introduce a constant $\pm 1/b^2$,
\begin{equation}
    \frac{f^2(r)\,{t'}^2(r)}{r^{2d}\left({\frac{1}{f(r)} - f(r){t'}^2(r)}\right)} 
    = \pm\frac{1}{b^2},
\end{equation}
where $b$ is a positive constant with dimensions of $[\text{length}]^d$, to be fixed by boundary conditions. The choice of sign distinguishes two qualitatively different classes of extremal surface solutions. Solving algebraically for $t'(r)$ gives the first-order differential equation (see also
\cite{Afrasiar:2025eam})
\begin{equation}
  t'(r)  = \pm \frac{ r^{d}}{f(r)\sqrt{r^{2d}  \pm b^2 f(r)} },
\end{equation}
with the boundary-reaching segments satisfying $t(\epsilon)=\pm T/2$, where $\epsilon$ is a UV cutoff. The overall $\pm$ sign labels the two time-reflected branches of the surface, while the $\pm$ inside the square root differentiates the two solution classes: the minus sign permits an exterior turning point, and the plus sign permits an interior turning point. In fact,
\[
\frac{1}{f(r)}-f(r)t'(r)^2
=\frac{\pm b^2}{r^{2d}\pm b^2f(r)}.
\]
Thus, wherever the profile is real, the minus-sign solution is timelike and the plus-sign solution is spacelike, including behind the horizon.

From the expression for $t'(r)$, a radial turning point $r_0$ is determined by the vanishing of the square root, $r_0^{2d}\pm b^2 f(r_0)=0$. The separate pole at $f(r)=0$ is a Schwarzschild coordinate singularity and does not by itself define a radial turning point. Depending on the sign choice, the square-root condition yields turning points either outside or inside the horizon:

\begin{itemize}
    \item {Exterior solution} (minus sign): The condition $r_0^{2d} - b^2 f(r_0) = 0$ with $f(r_0) > 0$ gives a turning point $r_0 < r_h$ outside the horizon.
    \item {Interior solution} (plus sign): The condition $r_0^{2d} + b^2 f(r_0) = 0$ with $f(r_0) < 0$ requires $r_0 > r_h$, placing the turning point inside the horizon.
\end{itemize}

\begin{figure}[h]
    \begin{center}
        \includegraphics[width=0.23\textwidth]{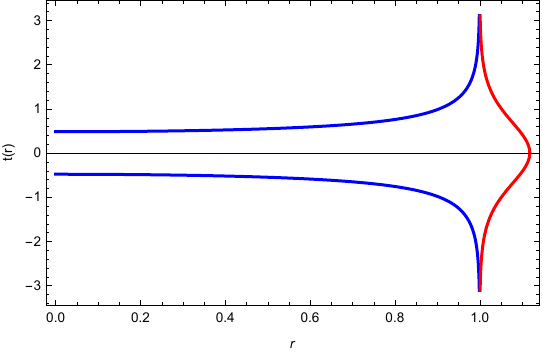}
        \includegraphics[width=0.23\textwidth]{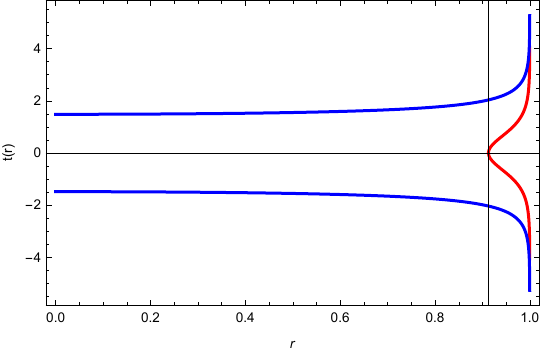}
        \caption{
Two examples of extremal surfaces. The left panel corresponds to the case where the turning point lies behind the horizon, while the right panel shows the configuration with two distinct branches. The blue curves represent the boundary-to-horizon segments and the red curves the horizon-to-turning-point segments. The red segment is timelike for an exterior turning point and spacelike for an interior turning point. The horizon is at $r_h=1$.
}
        \label{C1}
    \end{center}
\end{figure}

In both constructions considered below, the first, spacelike segment extends from the boundary cutoff $r=\epsilon$ to the horizon $r=r_h$, while the second segment connects the horizon to the turning point $r_0$. The second segment is timelike for an exterior turning point $r_0<r_h$ and spacelike for an interior turning point $r_0>r_h$. See figure \ref{C1}. The matching at the horizon used below cancels the common Schwarzschild coordinate pole; it does not by itself establish a smooth global junction. We therefore evaluate the regulated time and volume of these configurations without assuming that this matching supplies a complete extremal-surface prescription.

Interior horizon crossing also occurs for extremal surfaces in time-dependent backgrounds such as Vaidya geometries \cite{Liu:2013qca}. Here, however, the bulk geometry is static, and a timelike boundary interval does not introduce dynamical evolution of the background. The nontrivial radial profile instead describes the dependence of the configuration on the prescribed interval and must be supplemented by the appropriate global matching conditions.

Using the turning point condition, we can determine $b$ in terms of $r_0$:
\begin{equation}
b_1 = \frac{r_0^d}{\sqrt{f(r_0)}} \quad (\text{exterior}), \quad
b_2 = \frac{r_0^d}{\sqrt{-f(r_0)}} \quad (\text{interior}).
\end{equation}
The turning point $r_0$ is then related to the interval $T$ assigned by this construction via
\begin{equation}
T = -2\int_0^{r_0}t'(r)\,dr.
\end{equation}
Here the integration follows the appropriate radial segments through $r_h$, as displayed explicitly below; for an exterior turning point it is not an integral of a single real branch over $0<r<r_0$. The time integrals are written in the UV limit $\epsilon\to0$.

Before proceeding, we clarify the interpretation of $T$ and the structure of solutions. In the present setup, $T$ denotes the length of a timelike interval on the boundary, not a dynamical evolution time in a time-dependent background. Accordingly, we describe the different regimes in terms of small and large timelike intervals.

As the limits below show, the exterior interval vanishes both as $r_0\to0$ and as $r_0\to r_h$. Thus $T(r_0)$ is nonmonotonic, and a fixed interval need not determine a unique exterior turning point.

Explicitly,
\begin{align}
T = -2\bigg(&\int_0^{r_h}\frac{ r^{d}dr}{f(r)\sqrt{r^{2d} + b_1^2 f(r)} }\nonumber\\
      &+\int_{r_h}^{r_0} \frac{ r^{d}dr}{f(r)\sqrt{r^{2d} - b_1^2 f(r)} }\bigg) \quad (\text{exterior}),\nonumber \\
T =-2\bigg(&\int_0^{r_h}\frac{ r^{d}dr}{f(r)\sqrt{r^{2d} + b_2^2 f(r)} }\nonumber \\
      &+\int_{r_h}^{r_0} \frac{ r^{d}dr}{f(r)\sqrt{r^{2d} + b_2^2 f(r)} }\bigg) \quad (\text{interior}).
\end{align}

In both expressions, the first integral represents the contribution from the shared boundary-to-horizon branch. It is important to verify that this integral has no additional poles outside the horizon besides the expected divergence at $r=r_h$. The second integral, whose integrand differs in sign between the two configurations, captures the contribution from the horizon to the turning point. All horizon endpoints are regulated with a common radial cutoff, and the two contributions are combined before that cutoff is removed. For an interior turning point, equal offsets on the two sides of $r_h$ define the real principal value. The exterior expression contains only segments with $r\leq r_h$.

The timelike subregion complexity, following the CV prescription \eqref{complexity}, is computed from the volume of the bulk region enclosed by these extremal surfaces. After performing the transverse integration and integrating by parts in $r$, using
\[
\int\frac{t(r)}{r^{d+1}}\,dr
=-\frac{t(r)}{d r^d}
+\frac1d\int\frac{t'(r)}{r^d}\,dr,
\]
one obtains
\begin{align}\label{CC}
   \mathcal{C}_{\mathrm{T}}^{\mathrm{(ext)}} &= \frac{2L^{d-1}R^{d}}{dG}\bigg(\frac{T}{2\epsilon^d}
    + \int_{\epsilon}^{r_h}  \frac{ dr}{f(r)\sqrt{r^{2d} + b_1^2 f(r)} } \notag\\
    &\quad + \int^{r_0}_{r_h}  \frac{ dr}{f(r)\sqrt{r^{2d} - b_1^2 f(r)} }\bigg),\\
   \mathcal{C}_{\mathrm{T}}^{\mathrm{(int)}} &= \frac{2L^{d-1}R^{d}}{dG}\bigg(\frac{T}{2\epsilon^d}
    + \int_{\epsilon}^{r_h}  \frac{ dr}{f(r)\sqrt{r^{2d} + b_2^2 f(r)} } \notag\\
    &\quad +  \int^{r_0}_{r_h}  \frac{ dr}{f(r)\sqrt{r^{2d} + b_2^2 f(r)} }\bigg).
\end{align}
The factor of two accounts for time-reflection symmetry. The first term, $T/(2\epsilon^d)$, is the universal UV-divergent contribution proportional to the boundary time interval volume. The remaining integrals yield the finite part, with the same horizon prescription as for $T$. The volume is evaluated with the absolute value of the Lorentzian volume element and is real under this prescription. Imaginary contributions to the area functional are instead controlled by the induced signature, rather than horizon crossing alone.

The reality of the regulated volume does not by itself establish global admissibility or select a physical configuration at fixed $T$. We therefore regard the expressions above as regulated geometric volumes within the stated symmetric construction.

We now analyze the behavior of timelike complexity across different regimes, determined by the location of the turning point relative to the horizon: the deep exterior regime ($r_0 \ll r_h$), the near-horizon regime ($r_0 \simeq r_h$), and the interior penetration regime ($r_h < r_0 $). The quantities $\mathcal{C}_T$ are real under the stated principal-value and volume prescriptions, whose global interpretation remains subject to the matching conditions discussed above.

\subsection{Deep exterior regime}
When $r_0 \ll r_h$, the relevant extremal surface lies entirely in the exterior region, and the corresponding conserved quantity behaves as
\begin{equation}
b_1 \sim r_0^d .
\end{equation}
Although we assume $r_0 \ll r_h$, the radial coordinate in the first integral appearing in the expression for the boundary time interval $T$ ranges over the full interval $r \in [0, r_h]$. This individual integral has a logarithmic divergence near the horizon.
Setting
\begin{equation}
r = r_h (1 - \epsilon), \qquad \epsilon \to 0 ,
\end{equation}
where $\epsilon$ in this local parametrization is dimensionless and distinct from the UV cutoff, we have $f(r) \simeq (d+1)\epsilon$. Then,
the first integral contributing to $T$ behaves as
\begin{equation}
\int_0^{r_h}\frac{ r^{d}dr}{f(r)\sqrt{r^{2d} + b_1^2 f(r)} }
\;\sim\;
\int \frac{-r_h\, d\epsilon}{(d+1)\epsilon}
\;\sim\;
-\frac{r_h}{d+1}\ln\epsilon ,
\end{equation}
which is the familiar infrared logarithmic divergence near the horizon.

This divergence is canceled by the near-horizon contribution of the second integral in the expression for $T$. The leading finite term must nevertheless be obtained from both branches. The cancellation of their horizon poles does not justify discarding the finite contribution of the boundary-to-horizon segment.

In the region $r=\mathcal{O}(r_0)$ relevant to the leading small-$r_0$ term, we may approximate $f(r)\simeq1$, yielding the local expression
\begin{equation}
\frac{r^d}{f(r)\sqrt{r^{2d}-b_1^2 f(r)}}
\simeq\frac{r^d}{\sqrt{r^{2d}-r_0^{2d}}}.
\end{equation}
Changing variables to $u=r/r_0$ and retaining the matched difference of both branches, we find
\begin{align}
T={}&2r_0\bigg[\int_1^\infty u^d
\left(\frac1{\sqrt{u^{2d}-1}}
-\frac1{\sqrt{u^{2d}+1}}\right)du\nonumber\\
&\qquad-\int_0^1\frac{u^d\,du}{\sqrt{u^{2d}+1}}\bigg]+O(r_0)\\
={}&\frac{2\sqrt{\pi}\Gamma\left(\frac{d+1}{2d}\right)}
{\Gamma\left(\frac1{2d}\right)}
\tan\left(\frac{\pi}{4d}\right)r_0+O(r_0).
\end{align}
The difference in the first integral converges; the integral of either term separately would grow linearly at its upper limit.

Similarly, after combining the common horizon divergences, the small-$r_0$ contribution associated with the second volume integral is, for $d>1$,
\begin{align}
\int_{r_h}^{r_0}\frac{dr}{\sqrt{r^{2d}-r_0^{2d}}}
={}&-\frac{\sqrt{\pi}\Gamma\left(\frac{d-1}{2d}\right)}
{2d\,\Gamma\left(\frac{2d-1}{2d}\right)}
r_0^{1-d}\nonumber\\
&+O(r_0^{1-d}).
\end{align}
The first volume integral contributes at the same order. Thus, for $r_0\ll r_h$ and $d>1$, the finite part is
\begin{equation}
\begin{aligned}
\mathcal{C}_{\mathrm{T}}^{\mathrm{finite}}
={}&\frac{2L^{d-1}R^d}{dG}
\frac{\Gamma\left(\frac{2d+1}{2d}\right)
\Gamma\left(\frac{d-1}{2d}\right)}{\sqrt{\pi}}\\
&\times\left[1-\sin\left(\frac{\pi}{2d}\right)\right]
r_0^{1-d}+O(r_0^{1-d})\\
\sim{}&r_0^{1-d}\sim T^{1-d}.
\end{aligned}
\end{equation}
This indicates that on the short-interval branch approaching the boundary, the leading finite contribution agrees with the pure AdS strip limit.

We emphasize that this power-law behavior for $d>1$ is not directly comparable to the pure AdS hyperbolic results of Eqs.~\eqref{eq:exact}. Instead, it should be compared with the pure AdS strip entangling region in Poincar\'e coordinates, for which the timelike subregion complexity exhibits the same $T^{1-d}$ scaling in the finite part. This agreement concerns the leading small-$r_0$ contribution.

\subsection{Near-horizon regime}

We now consider turning points close to the horizon,
$r_0\simeq r_h$, approaching from either side.\footnote{I would
like to thank M.~Rahimi for pointing out errors in the near-horizon
computations of the original version and for discussions clarifying
the behavior in this regime.}
For the exterior approach, we write
\begin{equation}
r_0=r_h(1-\delta),\qquad \delta\to0^+,
\end{equation}
so that
\begin{equation}
b_1^2=\frac{r_0^{2d}}{f(r_0)}
\simeq\frac{r_h^{2d}}{(d+1)\delta}.
\end{equation}

The individual integrals contributing to $T$ and $\mathcal{C}_T$
contain logarithmic horizon divergences that cancel when the two
branches are combined with a common cutoff. Their finite parts
require matching the near-horizon contributions to the remainder
of the boundary-to-horizon integral.

On the short segment $r_0<r<r_h$, we have
$f(r)\simeq(d+1)(r_h-r)/r_h$. The singular part of the second
time integral is therefore
\begin{align}
&\int_{r_h}^{r_0}
\frac{r^d\,dr}{f(r)\sqrt{r^{2d}-b_1^2f(r)}}
\nonumber\\
&\quad\simeq\frac{r_h}{d+1}
\int_{r_h}^{r_0}
\frac{dr}{(r_h-r)\sqrt{(r-r_0)/(r_h-r_0)}}
\nonumber\\
&\quad=\frac{r_h}{d+1}
\left[
\ln\frac{\sqrt{r_h-r_0}+\sqrt{r-r_0}}
{\sqrt{r_h-r_0}-\sqrt{r-r_0}}
\right]_{r_h}^{r_0}.
\end{align}
The turning endpoint $r=r_0$ is integrable and produces no
logarithm; the divergence arises only at the horizon.

To evaluate the finite interval, place both horizon endpoints at
$r=r_h(1-\epsilon)$ and remove the common cutoff $\epsilon$ at
fixed $b_1$, before taking $b_1/r_h^d\to\infty$. Here $\epsilon$
is dimensionless and distinct from the boundary UV cutoff.
Changing the integration variable to $f(r)$, with
\[
r^d\,dr=-\frac{r_h^{d+1}}{d+1}\,df(r),
\]
gives the matched expansions
\begin{align}
&\int_0^{r_h(1-\epsilon)}
\frac{r^d\,dr}{f(r)\sqrt{r^{2d}+b_1^2f(r)}}
\nonumber\\
&\quad=\frac{r_h}{d+1}\biggl[
\log\frac{4r_h^{2d}}{(d+1)b_1^2\epsilon}
-\frac{2r_h^d}{b_1}
\nonumber\\
&\qquad\qquad
+\frac{2d\,r_h^{2d}}{(d+1)b_1^2}
+O\left(\frac{r_h^{3d}}{b_1^3}\right)\biggr],
\nonumber\\
&\int_{r_h(1-\epsilon)}^{r_0}
\frac{r^d\,dr}{f(r)\sqrt{r^{2d}-b_1^2f(r)}}
\nonumber\\
&\quad=\frac{r_h}{d+1}\biggl[
-\log\frac{4r_h^{2d}}{(d+1)b_1^2\epsilon}
\nonumber\\
&\qquad\qquad
+\frac{2d\,r_h^{2d}}{(d+1)b_1^2}
+O\left(\frac{r_h^{4d}}{b_1^4}\right)\biggr].
\label{eq:near_horizon_matched_expansions}
\end{align}
Terms vanishing with the horizon cutoff are omitted.
The logarithms cancel in the sum, and Eq.~(23) yields
\begin{equation}
T=\frac{4r_h^{d+1}}{(d+1)b_1}
-\frac{8d\,r_h^{2d+1}}{(d+1)^2b_1^2}
+O\left(\frac{r_h^{3d+1}}{b_1^3}\right).
\label{eq:near_horizon_time_b}
\end{equation}
Using the turning-point relation
$r_h^d/b_1=\sqrt{(d+1)\delta}+O(\delta^{3/2})$, we obtain
\begin{equation}
T=\frac{4r_h}{\sqrt{d+1}}\sqrt{\delta}
-\frac{8d\,r_h}{d+1}\delta
+O(r_h\delta^{3/2}).
\label{eq:near_horizon_time_delta}
\end{equation}
Inverting this expression gives
\begin{equation}
\delta=\frac{d+1}{16}\left(\frac{T}{r_h}\right)^2
+O\left(\frac{T^3}{r_h^3}\right).
\end{equation}
Thus the exterior near-horizon limit corresponds to $T\to0$.
Although the Hawking temperature $T_H=(d+1)/(4\pi r_h)$ sets
the thermal scale, the approach to the horizon is algebraic
in $T$, not exponential at long intervals.

The finite volume likewise requires both terms in \eqref{CC}.
In the volume integrals below, $\epsilon$ again denotes the
boundary UV cutoff, while the horizon endpoints are understood
with the common-cutoff prescription. Removing the UV cutoff
before taking the near-horizon limit gives
\begin{align}
&\lim_{\epsilon\to0}\biggl[
\int_\epsilon^{r_h}
\frac{dr}{f(r)\sqrt{r^{2d}+b_1^2f(r)}}
\nonumber\\
&\qquad\qquad
+\int_{r_h}^{r_0}
\frac{dr}{f(r)\sqrt{r^{2d}-b_1^2f(r)}}\biggr]
\nonumber\\
&\quad=\frac{r_h}{b_1}
\frac{(d-1)\sqrt{\pi}\,
\Gamma\left(1+\frac{1}{d+1}\right)}
{(d+1)\Gamma\left(\frac12+\frac{1}{d+1}\right)}
+O\left(\frac{r_h}{b_1}\right).
\end{align}
Consequently,
\begin{align}
\mathcal{C}_T^{\mathrm{finite}}
={}&\frac{2L^{d-1}R^d}{dG r_h^{d-1}}
\frac{(d-1)\sqrt{\pi}\,
\Gamma\left(1+\frac{1}{d+1}\right)}
{\sqrt{d+1}\,
\Gamma\left(\frac12+\frac{1}{d+1}\right)}
\sqrt{\delta}
\nonumber\\
&+O(\sqrt{\delta}).
\end{align}
Combining this result with
\eqref{eq:near_horizon_time_delta} gives the leading
short-interval relation
\begin{equation}
\begin{aligned}
\mathcal{C}_T^{\mathrm{finite}}
={}&\frac{4S}{dL}
\frac{(d-1)\sqrt{\pi}\,
\Gamma\left(1+\frac{1}{d+1}\right)}
{2\Gamma\left(\frac12+\frac{1}{d+1}\right)}\,T\\
&+O\left(\frac{ST}{L}\right),
\qquad T\to0,
\end{aligned}
\label{Clinear}
\end{equation}
where $S=R^dL^d/(4Gr_h^d)$ is the Bekenstein--Hawking entropy
of the black brane for spatial volume $L^d$.

For the interior approach, we instead write
\begin{equation}
r_0=r_h(1+\delta),\qquad \delta\to0^+,
\end{equation}
with
\begin{equation}
b_2^2=-\frac{r_0^{2d}}{f(r_0)}
\simeq\frac{r_h^{2d}}{(d+1)\delta}.
\end{equation}
On the segment $r_h<r<r_0$,
$f(r)\simeq-(d+1)(r-r_h)/r_h$. Equal radial offsets on the
two sides of the horizon define the real principal value.
Combining the two time integrals before removing this cutoff
gives
\begin{align}
T&=\frac{4r_h^{d+1}}{(d+1)b_2}
+O\left(\frac{r_h^{3d+1}}{b_2^3}\right)
\nonumber\\
&=\frac{4r_h}{\sqrt{d+1}}\sqrt{\delta}
+O(r_h\delta^{3/2}).
\end{align}
With the same order of limits, the paired finite-volume
integrals satisfy
\begin{align}
&\lim_{\epsilon\to0}\biggl[
\int_\epsilon^{r_h}
\frac{dr}{f(r)\sqrt{r^{2d}+b_2^2f(r)}}
\nonumber\\
&\qquad\qquad
+\int_{r_h}^{r_0}
\frac{dr}{f(r)\sqrt{r^{2d}+b_2^2f(r)}}\biggr]
\nonumber\\
&\quad=\frac{r_h}{b_2}
\frac{(d-1)\sqrt{\pi}\,
\Gamma\left(1+\frac{1}{d+1}\right)}
{(d+1)\Gamma\left(\frac12+\frac{1}{d+1}\right)}
+O\left(\frac{r_h}{b_2}\right).
\end{align}
The interior branch therefore obeys the same leading
finite-complexity relation \eqref{Clinear}.

Both approaches share the endpoint $T=0$ and
$\mathcal{C}_T^{\mathrm{finite}}=0$. Inverting their leading
time dependence gives
\begin{equation}
r_0=r_h\mp\frac{d+1}{16r_h}\,T^2
+O\left(\frac{T^2}{r_h}\right),
\end{equation}
where the upper and lower signs correspond to the exterior
and interior approaches, respectively. Thus $dr_0/dT\to0$
on both branches, and their finite-complexity curves have
the same leading tangent as $T\to0$. This agreement concerns
the limiting interval and finite volume; it does not make
$T(r_0)$ differentiable at the horizon, since
\begin{equation}
\frac{dT}{dr_0}\sim
\begin{cases}
-\dfrac{2}{\sqrt{(d+1)\delta}},
& r_0=r_h(1-\delta),\\[2mm]
+\dfrac{2}{\sqrt{(d+1)\delta}},
& r_0=r_h(1+\delta).
\end{cases}
\end{equation}
Nor does agreement of these limiting observables establish
a smooth global junction of the bulk surface segments.

The common slope in \eqref{Clinear} is determined by $S/L$
and a dimension-dependent factor. Since $T$ is the length
of the prescribed boundary interval at fixed geometry,
$d\mathcal{C}_T/dT$ describes interval dependence rather
than dynamical complexity growth. This short-interval
relation therefore provides no direct comparison with
Lloyd's bound.

\subsection{Interior Penetration and Maximum Depth}

We now consider the plus-sign solution with an interior turning point, \(r_0>r_h\), and ask how deeply it can penetrate behind the horizon. The turning point satisfies
\begin{equation}
V(r_0)=0,\qquad
V(r)\equiv r^{2d}+b_2^2 f(r).
\end{equation}
For \(r_0\) to be the first zero reached from the boundary while the radicand remains nonnegative, it must also satisfy \(V'(r_0)<0\). The limiting depth \(r_{\max}(d)\) is therefore determined by the double-root conditions
\begin{equation}
V(r_{\max})=0,\qquad
V'(r_{\max})=0.
\end{equation}
Using
\begin{equation}
f'(r)=-\frac{d+1}{r_h^{d+1}},r^d,
\end{equation}
we obtain
\begin{equation}\label{rmax}
r_{\max}
=r_h\left(\frac{2d}{d-1}\right)^{\frac{1}{d+1}},
\qquad d>1.
\end{equation}
Beyond \(r_{\max}\), a zero of \(V(r)\) cannot be the first boundary-accessible turning point of this real family, since the radicand becomes negative before that zero is reached. Thus, \eqref{rmax} bounds the turning points of the profiles considered here, rather than establishing a universal causal bound on extremal surfaces. Moreover, \(r_{\max}\to r_h^+\) as \(d\to\infty\), so the allowed interior range contracts toward the horizon.

For
\begin{equation}
r_h<r_0<r_{\max}(d),
\end{equation}
the horizon pole is treated using the prescribed principal-value prescription. Because the independent scale \(r_h\) remains, dimensional analysis alone does not imply \(T\sim r_0\) throughout this range. Instead, expanding around the double root yields, as \(r_0\to r_{\max}^-\),
\begin{equation}
\begin{aligned}
T={}&\frac{2r_{\max}}{d+1}\sqrt{\frac{d-1}{d}}
\ln\left(\frac{r_h}{r_{\max}-r_0}\right)
+\mathcal{O}(r_h),\\
\mathcal{C}_T^{\mathrm{finite}}
={}&-\frac{L^{d-1}R^d}{dG r_{\max}^{d}}\,T
+\mathcal{O}\left(\frac{L^{d-1}R^d}{G r_h^{d-1}}\right).
\end{aligned}
\end{equation}
The logarithmic divergence arises from the coalescing roots of \(V(r)\), not from the horizon pole. The negative finite remainder does not imply that the complete UV-divergent volume is negative.

The scale \(r_{\max}\) thus marks the limiting first root of this real interior family and is approached only as \(T\to\infty\). It characterizes the relation between interval size and accessible turning-point depth within this construction, but does not, by itself, establish a broader interpretation in terms of boundary temporal resolution or causal access. In particular, referring to \(r_{\max}\) as a universal ``complexity horizon'' would require additional evidence, since the branch is spacelike and its global boundary interpretation remains unsettled.

These results distinguish the short-interval exterior limits from the logarithmic interior critical limit. Although the regulated volume is real under the stated prescription, reality alone does not establish a complete holographic observable: the global joining of surface segments and the selection of admissible configurations at fixed boundary data must also be specified.

Finally, these static interval dependences should not be confused with complexity growth under time evolution \cite{Susskind:2014rva, Stanford:2014jda}. In particular, the near-horizon relation \eqref{Clinear} describes a small-interval limit, not the dynamical growth rate appearing in Lloyd's bound \cite{Lloyd:2000Nature}.


\section{Conclusion and Outlook}

In this work, we have investigated timelike subregion complexity within the holographic Complexity=Volume (CV) framework. Motivated by the pseudo-entropy program, we have studied the bulk volume bounded by extremal surfaces associated with finite boundary time intervals of length $T$. Importantly, $T$ denotes the interval length, not the evolution time of a fixed boundary configuration.

For timelike hyperbolic regions in pure AdS$_{d+2}$, the prescribed volume bounded by spacelike and timelike extremal surfaces remains real. The same holds for thermal states dual to AdS black branes, where extremal surfaces can remain outside the horizon or penetrate the black hole interior. Unlike pseudo-entropy, which may acquire imaginary contributions under analytic continuation, the reality of $\mathcal{C}_T$ follows from its definition as a regulated Lorentzian bulk volume. Its precise boundary interpretation, however, remains open.

The existence of exterior and interior extremal surface branches raises questions about global surface selection and possible transitions, reminiscent of those encountered in spacelike entanglement entropy in eternal black holes \cite{Hartman:2013qma}. In our construction, the exterior relation $T(r_0)$ is nonmonotonic, so a fixed interval length need not determine a unique turning point. The exterior and interior branches share the limit $T=0$ and the same leading finite-complexity slope near the horizon. This agreement does not, by itself, establish a smooth continuation between complete surfaces or determine which configuration should be selected. Such questions require global matching and admissibility conditions beyond the turning-point equations.

A central geometric result is a dimension-dependent upper bound on the first accessible turning point of the real interior branch:
\begin{equation}
\frac{r_{\max}(d)}{r_h}
=
\left(\frac{2d}{d-1}\right)^{\frac{1}{d+1}},
\qquad d>1.
\end{equation}
This bound arises when the extremal surface equation develops a double root and is approached as $T\to\infty$. Beyond $r_{\max}(d)$, a zero of the radicand cannot serve as the first boundary-accessible turning point of this family. The result establishes a classical geometric obstruction within our prescription, but interpreting it as a universal causal bound or an intrinsic complexity horizon requires further justification.

The interval dependence of $\mathcal{C}_T$ exhibits several distinct regimes. For small intervals on the exterior branch, with $r_0\ll r_h$ and $d>1$, the finite part scales as
\begin{equation}
  \mathcal{C}_T^{\mathrm{finite}}\sim T^{1-d},  
\end{equation}
reproducing the leading pure AdS strip behavior. Near the horizon, both exterior and interior branches exhibit the same leading linear relation between $\mathcal{C}_T^{\mathrm{finite}}$ and $T$, with the dimension-dependent coefficient given in \eqref{Clinear}. This relation does not extend across the entire interior branch. For large interior intervals, as $r_0\to r_{\max}$, the finite part instead behaves as
\begin{equation}
    \mathcal{C}_T^{\mathrm{finite}}
\sim
-\frac{L^{d-1}R^d}{dG\,r_{\max}^d}\,T.
\end{equation}
The interval length diverges logarithmically as the limiting turning point is approached.

These relations describe dependence on interval length, not dynamical complexity growth. In particular, $d\mathcal{C}_T/dT$ is not a growth rate for the time evolution of a fixed state and cannot be directly compared with Lloyd's bound \cite{Lloyd:2000Nature}. Nor does the finite range of interior turning points imply a finite complexity budget, since $T$ diverges as $r_0\to r_{\max}^-$.

A natural direction for future work is to establish a boundary interpretation of timelike subregion complexity. Possible approaches include circuit complexity, path-integral optimization, and Lorentzian tensor networks. Any circuit interpretation must specify the allowed operations, reference data, and cost function. In particular, our geometric construction does not establish a minimal circuit depth for transforming $\rho_A(t_i)$ into $\rho_A(t_f)$ using operations confined to $A$, since reduced-state evolution need not be unitary within the subregion.

The penetration bound also raises questions about bulk interior reconstruction. The scale $r_{\max}(d)$ limits the first turning point of the interior family considered here. Relating this bound to entanglement wedge reconstruction requires a complete surface prescription and a clear boundary interpretation of the regulated volume. The double-root obstruction alone does not establish access to regions inaccessible to entanglement-based observables.

Several further extensions are worth exploring. A comparison with the Complexity=Action (CA) proposal could determine whether the reality of timelike complexity persists under a different bulk prescription. Time-dependent geometries, including Vaidya backgrounds and evaporating black holes, could reveal how the penetration bound changes dynamically. Adding charge, rotation, or quantum extremal surfaces would further test its robustness and possible quantum-gravitational corrections.

Finally, recent studies of spatial subregion complexity \cite{Fan:2025moc,Haah:2025hyf} have found sharp dynamical transitions depending on subsystem size and, in holographic models, temperature. Unlike those studies, our construction varies the length of a timelike interval in a static setting. Investigating whether a complete prescription exhibits transitions between admissible extremal surface branches could reveal useful connections while preserving the distinction between interval dependence and time evolution.

\section*{Acknowledgements}

I would like to thank Andreas Karch, Souvik Banerjee and Tadashi Takayanagi for their valuable comments. I am also grateful to Kyriakos Papadodimas and Mohammad Javad Vasli for many insightful discussions on various aspects of holographic complexity. I would further like to thank the CERN Department of Theoretical Physics for their warm hospitality during the course of this work.
This research was supported by the Iran National Science Foundation (INSF) under Project No.~4023620. I also acknowledge the assistance of ChatGPT for editorial help in refining and polishing the manuscript.

\bibliographystyle{apsrev4-2}

\end{document}